\documentclass[preprint,prl,floatfix,superscriptaddress]{revtex4}
\usepackage{epsfig}
\usepackage{amsmath,amssymb}
\usepackage{bm}
\usepackage{color,colordvi,graphicx}

\begin{document}
\title{Electromagnetically induced spatial light modulation}

\author{L. Zhao}
\affiliation{Department of Physics, University of Connecticut,
Storrs, Connecticut 06269}

\author{T. Wang}
\affiliation{Department of Physics, University of Connecticut,
Storrs, Connecticut 06269}

\author{S. F. Yelin}
\affiliation{Department of Physics, University of Connecticut,
Storrs, Connecticut 06269} \affiliation{ITAMP, Harvard-Smithsonian Center
for Astrophysics, Cambridge, Massachusetts 02138}

\date{\today}
\begin{abstract}We theoretically report that, utilizing electromagnetically induced transparency (EIT), the transverse spatial properties of weak probe fields can be fast modulated by using optical patterns (e.g. images) with desired intensity distributions in the coupling fields. Consequently, EIT systems can function as high-speed optically addressed spatial light modulators. To exemplify our proposal, we indicate the generation and manipulation of Laguerre-Gaussian beams based on either phase or amplitude modulation in hot vapor EIT systems.
\end{abstract}

\maketitle 

\noindent Spatial light modulator (SLM) technology plays a significant role in modern optics \cite{slm}. This technology can impose phase and amplitude modulations on light fields in real time, thus performing spatial information encoding and producing light fields with novel spatial structures. In optics, Laguerre-Gaussian (LG) modes with isolated intensity minima in their transverse pofiles have attracted a great deal of interest \cite{allen,oam5}. Numerous reseach has shown that LG modes carry well-defined orbital angular momentum which can impart on matter through light-matter interaction or characterize multidimensional quantum states of light in Hilbert space. Therefore, LG beams have promising applications in optical tweezers and traps \cite{oam1,oam2}, dense coding in quantum information processing \cite{oam4,oam6}. etc. To generate and manipulate LG beams, programmable liquid crystal SLMs are increasingly adopted \cite{oam1,oam2,oam4,oam5}. Limited by the inertia of liquid crystal molecules, the refresh rates of liquid crystal SLMs are at most a few kHz. However, for many applications based on LG beams, such as dynamic optical traps and large capacity  parallel information processing, much faster reshaping of light fields is expected. Thus, liquid crystal SLMs could pose a real difficulty to the applications. 

\begin{figure}[ht]
\centerline{\includegraphics[clip,angle=0,width=0.85\linewidth]{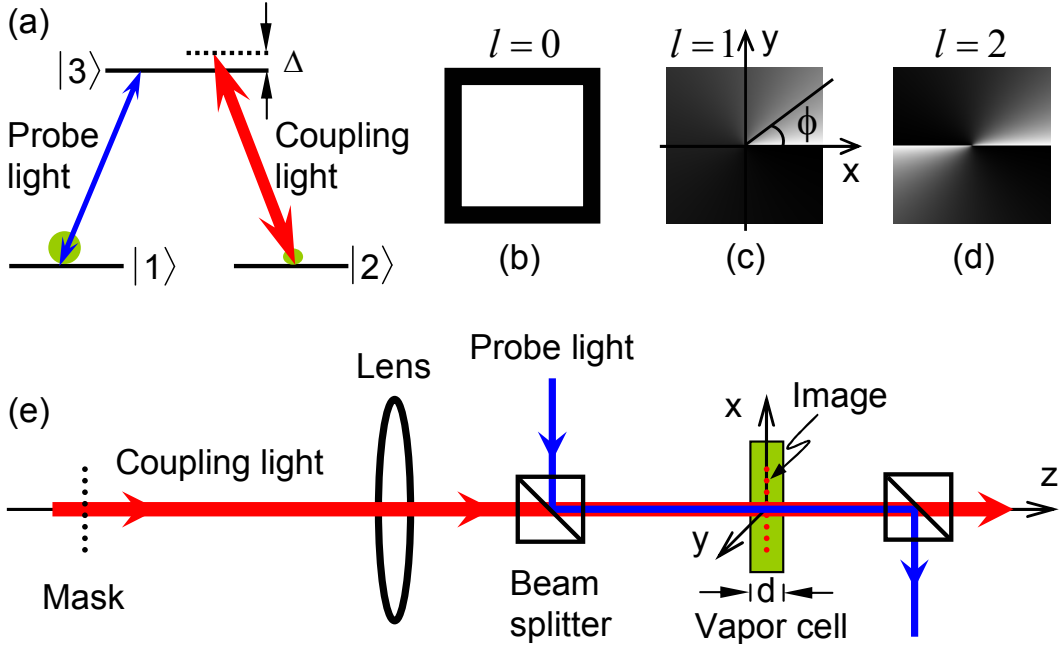}}
\caption{(Color online) (a) Three-level $\Lambda$ atomic system and two copropagating light fields with a small detuning $\Delta$. (b) Uniform illumination of the coupling field imposing no phase shift onto the incident Gaussion probe field ($l=0$). (c) Azimuthal intensity distribution of the coupling field to generate the LG beam with $l=1$,  where $\phi$ is the azimuthal angle. The intensity distribution could be shown numerically in Fig. \ref{fig:xp}(a). (d) $l=2$. (e) Possible experimental setup which includes an imaging system, a vapor cell of the thickness $d$ and some beam splitters.}
\label{fig:eit1}
\end{figure}

Electromagnetically induced transparency (EIT), as a phenomenon of quantum interference, has been intensively studied for decades \cite{eit}. In EIT systems, the dispersion and absorption of weak probe light fields can be coherently controlled by strong coupling light fields via atomic transitions. Moreover, the atomic response time to the light fields could be much shorter than that of the liquid crystal molecules \cite{speed1, speed2}. Consequently, the phase and amplitude of the probe fields could be modulated all-optically and quickly in the EIT media. Previously, EIT systems were widely used to manipulate light fields without transverse structures \cite{lukin}, while some recent studies have focused on processing multimode transverse images in EIT systems, particularly, in low-cost vapor cells \cite{mip}. Therefore, EIT systems might also be a good candidate to perform spatial light modulation with high speed.

In this paper we examine the possibility of spatially modulating probe light beams in EIT systems with the use of optical patterns (e.g. images) with desired intensity distributions in the strong coupling fields. As an example, the generation and manipulation of LG beams are investigated. In our scheme, we consider a three-level $\Lambda$ atomic system interacting with two copropagating light fields in a hot vapor cell (Fig. \ref{fig:eit1}), which can strongly suppress the Doppler broadening. The weak probe field of frequency $\omega_{p}$ and amplitude $E_{p}$ is resonant with the transition $\vert 1 \rangle \leftrightarrow \vert 3 \rangle$ ($\omega_{31}=\omega_{p}$). The strong coupling field of frequency $\omega_{c}$ and amplitude $E_{c}$ drives the transition $\vert 2 \rangle \leftrightarrow \vert 3 \rangle$ with a small single-photon detuning $\Delta=\omega_{32}-\omega_{c}$. The Rabi frequencies of the probe and coupling fields are defined as $\Omega_{p,c}=\mu_{13,23} \cdot E_{p,c}/\hbar$, where $\mu_{13,23}$ are the dipole moments of the transitions and $\Omega_{p} \ll \Omega_{c}$ for the EIT situation. Thus, ignoring the Doppler broadening, the linear susceptibility for the probe field can be given by \cite{eit}

\begin{eqnarray} \label{sus}
& &\chi = \chi^{\prime} + i \chi^{\prime \prime} = {\vert \mu_{13} \vert}^{2} \varrho \times  \\
& &\frac{-4 \Delta {\vert \Omega_{c} \vert}^{2}+i 8 \Delta^{2} \gamma_{31}+i 2 \gamma_{21} ({\vert \Omega_{c} \vert}^{2}+\gamma_{21} \gamma_{31})}{\epsilon_{0} \hbar {\vert {\vert \Omega_{c} \vert}^{2}+\gamma_{31} (\gamma_{21}-i 2 \Delta) \vert}^{2}},   \nonumber 
\end{eqnarray}
where $\varrho$ is the the atom number density, $\gamma_{31}$ is the decay rate from $\vert 3 \rangle$ to $\vert 1 \rangle$ and $\gamma_{21}$ is the decoherence rate between  $\vert 1 \rangle$ and $\vert 2 \rangle$.

\begin{figure}[ht]
\centerline{ \includegraphics[clip,width=0.9\linewidth]{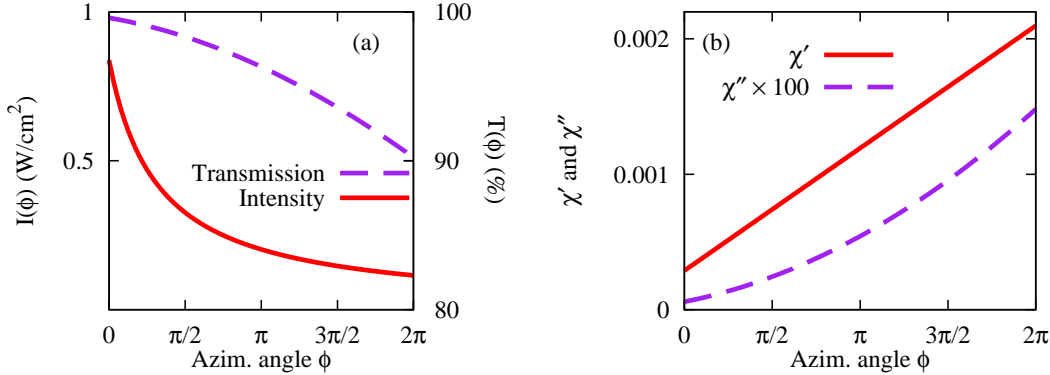}}
\caption{(Color online) (a) Azimuthal intensity distribution $I(\phi)$ ($=\epsilon_{0} c E_{c}(\phi)^{2}/2$) of the coupling field (solid) corresponding to Fig. \ref{fig:eit1}(c) with $a=500$, $b=c=1$ in Eq. (\ref{rabic}), where $I(0) = 838$ mW/cm$^{2}$ and $I(2\pi) = 115$ mW/cm$^{2}$. According to Eqs. (\ref{sus}), (\ref{rabic}) and the strong-coupling-field condition, this azimuthal intensity distribution of the coupling field can induce an azimuthally dependent refractive index and a small absorption for the probe field. Thus, also shown is the high transmission probability $T(\phi) > 90 \%$ of the probe field (dashed). (b) Linear susceptibility for the probe field in $^{87}$Rb induced by the azimuthal intensity distrbution in (a). The real part $\chi^{\prime}$ (solid) is associated with the azimuthally dependent refractive index and the imaginary part $\chi^{\prime \prime}$ (dashed) is associated with the small absorption.}
  \label{fig:xp}
\end{figure}

To generate LG modes by means of phase modulation, an azimuthal phase winding $e^{i l \phi}$ should be imprinted onto the wave front of the Gaussian probe field, where $l$ is the integer winding number and $\phi$ is the azimuthal angle \cite{oam1}. This phase imprinting requires an azimuthal variation of the refractive index $n(\phi)$ in the EIT system, which can be produced by an azimuthally dependent Rabi frequency $\Omega_{c}(\phi)$ of the coupling field according to Eq. (\ref{sus}). Intuitively, this coupling field can be constructed by the images of amplitude masks in an imaging system [Fig. \ref{fig:eit1}(b)-(e)]. To numerically evaluate this scheme, three sublevels in the $^{87}$Rb D2 line ($5^{2}S_{1/2} \leftrightarrow 5^{2}P_{3/2}$, $\lambda=780$ nm) are adopted, which can reasonably give $\gamma_{31}=38.11 \times 10^{6}/$s, $\mu_{13}=\mu_{23}=3.58 \times 10^{-29}$ C$\cdot$m, $\gamma_{21}=2\pi \times 3000/$s, $\Delta=-0.2 \gamma_{31}$, $\varrho=5 \times 10^{12}/$cm$^{3}$ \cite{online}. From Eq. (\ref{sus}), the Rabi frequency of the coupling field can have the form 
\begin{eqnarray} \label{rabic}
\Omega_{c}(\phi) = \frac{\mu_{23} \cdot E_{c}(\phi)}{\hbar} = \sqrt{\frac{a}{b \phi + c}} \gamma_{31},
\end{eqnarray}
to introduce the azimuthal dependence of $\chi$, where $a$, $b$, and $c$ are the adjustable parameters. When the strong-coupling-field condition $\Omega_{c}(\phi) \gg \{ \gamma_{31}, \Delta \} \gg \gamma_{21}$ 
are satisfied, Eqs. (\ref{sus}) and (\ref{rabic}) give $\chi^{\prime} \propto -\Delta/\Omega_{c}^{2} \propto \phi$ and $\chi^{\prime \prime} \propto \gamma_{21}/\Omega_{c}^{2} \rightarrow 0$ which tell us that the refractive index can vary almost linearly around the optical axis [the $z$ axis in Fig. \ref{fig:eit1}(e)] and the absorption can simultaneously be suppressed. For simplicity, we discuss here the generation of LG mode with $l=1$ from the Gaussion probe field with $l=0$. A $2\pi$ phase difference between the maximum and minimum should be imprinted onto the wave front of the probe field, which can result in
\begin{eqnarray} \label{2pi}
 \Delta n  \cdot d= [n(2\pi)-n(0)] \cdot d = \Delta l \cdot \lambda ,
\end{eqnarray}
where $n$ is the refractive index, $d$ is the thickness of the vapor cell and $\Delta l = 1$. It is known that, in EIT systems, $\chi^{\prime}$ and $\chi^{\prime \prime}$ are much smaller than unity ($\chi^{\prime}$, $\chi^{\prime \prime}$ $\ll 1$), which can lead to $n=\sqrt{1+\chi} \approx 1 + \chi^{\prime}/2 + i \chi^{\prime \prime}/2$. Thus, Eq. (\ref{2pi}) can be rewritten as
$[\chi^{\prime}(2\pi)-\chi^{\prime}(0)] \cdot d = 2 \lambda$.
Moreover, the transmission probability can read $ T=e^{-\alpha d}$, where the absorption coefficient $\alpha=2 \pi \chi^{\prime \prime}/ \lambda$. Assuming $a=500$, $b=c=1$ for the strong-coupling-field condition which can give $\Omega_{c}(\phi) \approx 10 \gamma_{31}$, and with the parameters in $^{87}$Rb, one can find the thickness of the vapor cell $d \approx 862 \ \mu$m and other features shown in Fig. \ref{fig:xp}. According to Eq. (\ref{2pi}), the $2 \pi$ phase modulation can thus be reached in a very thin cell, which is crucial to our scheme, because (i) the thin cell can lead to high resolution images through the EIT system and the alignment between the probe field and the images, (ii) it can further reduce the absorption of the system [$T > 90\%$ in Fig. \ref{fig:xp}(a)], (iii) it can ensure the validity of the paraxial approximation and nearly phase-only operations \cite{oam1, spp}. Additionally, high order LG modes can be created by more complex images, such as in Fig. \ref{fig:eit1}(d).

The switching between different LG modes (e.g. between $l=0$ and $l=1$) is essentially associated with the conversion between different dark states in the EIT system. The switching time depends on the following factors. The first one is the time scale for the re-establishment of dark states, which can be given by $\tau_{r} \approx \alpha d \gamma_{31}/\Omega_{c}^{2}$ \cite{speed1, speed2}. Suppose that we have the initial image in Fig. \ref{fig:eit1}(c) with the intensity distribution numerically shown in Fig. \ref{fig:xp}(a), then switch this image to a uniform illumination [Fig. \ref{fig:eit1}(b)] with $I_{uni}=I(0) = 838$ mW/cm$^{2}$. It can be seen that the weakest coupling intensity occurs at $\phi=2 \pi$ where the system has the highest absorption and, accordingly, the longest re-establishment time. We can find $\tau_{r} \approx \alpha d \gamma_{31}/\Omega_{uni}^{2} \approx 34$ ps at $\phi=2 \pi$, where $\Omega_{uni}$ is the Rabi frequency of the uniform illumination. Secondly, an adiabatic manipulation should be performed to keep the coherence of the system. The time scale can be given by $\tau_{a} = \gamma_{31}/\Omega_{uni}^{2} \approx 0.33$ ns \cite{eit, adi}. Hence, the limit for the switching time is $\tau_{s} \gg \tau_{r}$, $\tau_{a}$, which shows the possibility of modulation at up to a fraction of GHz.

\begin{figure}[ht] 
\centerline{ \includegraphics[clip,angle=0,width=0.5\linewidth]{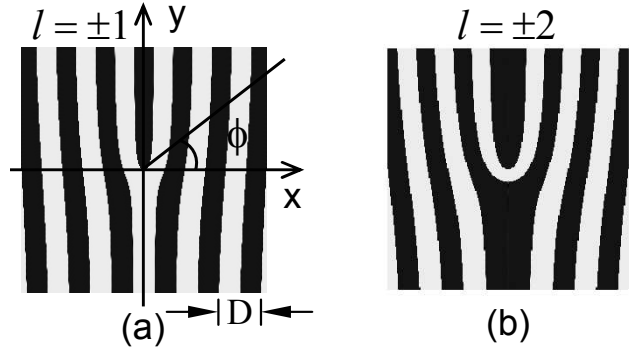}}
         \caption{(a) Forked binary amplitude grating to generate the LG modes with
$l= \pm 1$, where $\phi$ is the azimuthal angle and $D$ is the period of the grating far away from the forked center. (b) Similar grating for $l= \pm 2$.}
\label{fig:am}
\end{figure}

Next, we consider the generation of LG modes by means of amplitude modulation. In this case, the aforementioned images with azimuthal intensity distributions should be replaced by images with forked binary patterns (Fig. \ref{fig:am}) \cite{fork1,fork2}. Furthermore, we use the resonant three-level EIT system without detunings ($\Delta=0$). In the bright fringes of the forked coupling field, the probe field can transmit with very low absorption under the EIT condition ($T \rightarrow 1$). Nevertheless, in the dark fringes without the coupling intensity ($\Omega_{c}=0$), the probe field resonantly drives the transition $\vert 1 \rangle \leftrightarrow  \vert 3 \rangle$, which can result in a strong absorption. From Eq. (\ref{sus}) with $\Delta=\Omega_{c}=0$, we can obtain
\begin{eqnarray} \label{abs}
 \chi^{\prime \prime} = \frac{2 \mu_{12}^{2} \varrho}{\epsilon_{0} \hbar 
\gamma_{31}} \approx 0.0072,  \Rightarrow  T=e^{-\alpha d} \approx e^{-29} \rightarrow 0, 
\end{eqnarray}
where we use $d = 500 \ \mu$m for the thin cell condition and $\varrho = 10^{11}$ cm$^{-3}$, and preserve all the other parameters in $^{87}$Rb. Thus, a tunable forked binary amplitude grating is formed in the EIT system for the probe field. The transmission function of this grating can be given by \cite{fork2}
\begin{eqnarray} \label{tf}
t_{f}(r, \phi)=\frac{1}{2}+\frac{2}{\pi} \sum_{k=0}^{\infty} \frac{1}{2k+1} \times \nonumber \\
\sin[(2k+1)(-l \phi + \frac{2 \pi}{D} r \cos \phi)], 
\end{eqnarray}
where $ r =\sqrt{x^{2}+y^{2}}$, $\phi$ is the azimuthal angle, $l$ is the winding number and $D$ is the period of the grating far away from the forked center. The Fraunhofer diffraction pattern can be analytically derived and show that, the beams diffracted into the first order maxima of the pattern can be converted into the LG modes with the $\pm l$ winding numbers \cite{fork2}.

To switch the LG modes (e.g. between $l=0$ and $l=1$), we can switch the coupling fields between the uniform illumination [Fig. \ref{fig:eit1}(b)] and the forked grating [Fig. \ref{fig:am}(a)]. At the positions of the dark fringes, there should be a switching between the two-level resonant system and the three-level EIT system. Using the same methods mentioned above \cite{speed1,speed2,eit,adi} and assuming that the bright fringes and the uniform illumination have the same intensity (e.g. $I_{bf}=I_{uni}=838$  mW/cm$^{2}$), we can calculate the re-establishment time $\tau_{r} \approx 9.56$ ns and the time scale for the adiabatic approximation $\tau_{a} \approx 0.33$ ns. The limit for the switching time ($\tau_{s} \gg \tau_{r}$, $\tau_{a}$) shows that the modulation can be performed at up to many MHz.

In summary, we have demonstrated that LG modes can be generated and manipulated in low-cost hot vapor EIT systems with high speed. Since $2 \pi$ phase modulation with low absorption or binary amplitude modulation can be easily obtained in EIT systems by using images with the desired intensity distributions in the coupling fields, EIT systems can be used as high-speed low-cost optically addressed SLMs \cite{slm}. By employing more complex images, one can achieve light fields with more peculiar spatial structures \cite{mov}. Our scheme provides a flexible and practical way to realize the high-speed spatial light modulation in EIT systems  with potential applications in optical technology and information science.

We would like to acknowledge funding from NSF. L. Z. thanks J. Javanenien, Y. Xiao, L. Fang, R. Zhou and F. Peng for helpful discussions.


\begin{thebibliography}{99}

\bibitem{slm} U. Efron, editor. {\it Spatial Light Modulator Technology.}
  (Marcel Dekker, New York, 1994).

\bibitem{allen} L. Allen, M. W. Beijersbergen, R. J. C. Spreeuw, and J. P. Woerdman, Phys. Rev. A {\bf 45}, 8185 (1992)


\bibitem{oam5} G. Molina-Terriza, J. P. Torres, and L. Torner, Nature Phys.  {\bf 3}, 305 (2007).

\bibitem{oam1} D. G. Grier, Nature (London) {\bf 424}, 810 (2003).


\bibitem{oam2} F. K. Fatemi and M. Bashkansky, Opt. Express {\bf 14}, 1368 (2006).

\bibitem{oam4} A. Mair, A. Vaziri, G. Weihs, and A. Zeilinger, Nature (London) {\bf 412}, 313 (2001).

\bibitem{oam6} J. T. Barreiro, T.-C. Wei, and P. G. Kwiat, Nature Phys. {\bf 4}, 282 (2008).

\bibitem{eit}
M. Fleischhauer, A. Imamoglu, and J. P. Marangos, Rev. Mod. Phys. {\bf 77}, 633 (2005).

\bibitem{speed1}
H. Schmidt and R. J. Ram, Appl. Phys. Lett. {\bf 76}, 3173 (2000).

\bibitem{speed2} S. E. Harris and L. F. Luo, Phys. Rev. A {\bf 52}, R928 (1995).

\bibitem{lukin} M. D. Lukin, Rev. Mod. Phys. {\bf 75}, 457 (2003).

\bibitem{mip} L. Zhao, T. Wang, Y. Xiao and S. F. Yelin, Phys. Rev. A, {\bf
  77}, 041802(R) (2008); T. Wang, L. Zhao, L. Jiang, and S. F. Yelin, Phys. Rev. A 
{\bf 77}, 043815 (2008); P. K. Vudyasetu, R. M. Camacho, and J. C. Howell, Phys. Rev. Lett. {\bf 100}, 123903 (2008); M. Shuker, O. Firstenberg, R. Pugatch, A. Ron, and N. Davidson, Phys. Rev. Lett. {\bf 100}, 223601 (2008).


\bibitem{online} Daniel A. Steck, “Rubidium 87 D Line Data,” available
  online at http://steck.us/alkalidata (revision 2.1, 1 September 2008).

\bibitem{spp}
M. W. Beijersbergen, R. P. C. Coerwinkel, M. Kristensen, and J. P. Woerdman, Opt. Commun.
{\bf 112}, 321 (1994).

\bibitem{adi}
M. Fleischhauer and A. S. Manka, Phys. Rev. A {\bf 54}, 794 (1996).

\bibitem{fork1} N. R. Heckenberg, R. McDuff, C. P. Smith, H. Rubinsztein-Dunlop, and M. J. Wegener, Opt. Quantum Electron. {\bf 24}, S951 (1992).

\bibitem{fork2} G. F. Brand, J. Mod. Opt. {\bf 44}(6), 1243 (1997); Am. J. Phys. {\bf 67}(1), 55 (1999).

\bibitem{mov} J. E. Curtis and D. G. Grier, Opt. Lett. {\bf 28}(11), 872 (2003).

\end{thebibliography}
\end{document}